\documentclass[journal, compsoc, onecolumn]{IEEEtran}

\usepackage{todonotes}

\usepackage{titlesec}

\usepackage[utf8]{inputenc}
\usepackage{braket}
\usepackage{amsmath}

\usepackage{graphicx}
\usepackage{dcolumn}
\usepackage{bm}
\usepackage{hyperref}

\usepackage{mathrsfs}
\usepackage{relsize}

\usepackage{flushend}

\usepackage{listings}

\begin{document}

\title{Efficient Quantum Circuit Design with a Standard Cell Approach, with an Application to Neutral Atom Quantum Computers}

\author{
\IEEEauthorblockN{Evan Dobbs}
\IEEEauthorblockA{Aalto University, Espoo, Finland, 
evan.dobbs@aalto.fi\\}
\and
\IEEEauthorblockN{Joseph S. Friedman}
\IEEEauthorblockA{University of Texas at Dallas, Richardson, USA, 
joseph.friedman@utdallas.edu\\}
\and
\IEEEauthorblockN{Alexandru Paler}
\IEEEauthorblockA{Aalto University, Espoo, Finland, 
alexandru.paler@aalto.fi\\}
}

\maketitle

\begin{abstract}
We design quantum circuits by using the standard cell approach borrowed from classical circuit design, which can speed-up the layout of circuits with a regular structure. Our standard cells are general and can be used for all types of quantum circuits: error-corrected or not. The standard cell approach enables the formulation of layout-aware routing algorithms. Our method is directly applicable to neutral atom quantum computers supporting qubit shuttling. Such computers enable zoned architectures for memory, processing and measurement, and we design circuits using qubit storages (memory and measurement zones) and standard cells (processing zones). Herein, we use cubic standard cells for Toffoli gates and, starting from a 3D architecture, we design a multiplication circuit. We present evidence that, when compared with automatic routing methods, our layout-aware routers are significantly faster and achieve shallower 3D circuits (by at least 2.5x) and with a lower routing cost. Additionally, our co-design approach can be used to estimate the resources necessary for a quantum computation without using complex compilation methods. We conclude that standard cells, with the support of layout-aware routing, pave the way to very large scale methods for quantum circuit compilation.
\end{abstract}

\IEEEpeerreviewmaketitle

\section{Introduction}
\label{sec:intro}

The optimal design of large-scale circuits -- both quantum and classical -- is not computationally tractable, necessitating the use of sub-optimal heuristics. Noting that the qubit layout of a quantum computer is generally regular and similar to a tiling, large-scale quantum circuit design challenges can be naturally mapped to classical circuit design heuristics based on standard cells. We therefore propose the development of automated quantum circuit design algorithms that leverage the enormous repository developed through several decades of classical circuit design.

This paper is organised as follows: Section~\ref{sec:motiv} introduces the goals of our approach, and Section~\ref{sec:rel} connects and differentiates our method from the state-of-the-art. Section~\ref{sec:na} describes the natural application of our approach to the scalable compilation of neutral atom quantum circuits. Section~\ref{sec:method} is describing the standard cells, and the details about how these are used for compilation (Section~\ref{sec:compiling}). Section~\ref{sec:routing} shows how the tiles are helping and speeding up the routing procedures. Section~\ref{sec:lowdepth} will illustrate by numbers the advantage of the tiles we are using in this paper. Finally, we compare tiling with automatic compilation and the results are presented in Section~\ref{sec:res}.

\subsection{Motivation}
\label{sec:motiv}

This work is motivated by the need to compile efficient circuits as quickly as possible. Compilation speed is advantageous when estimating the resources necessary to run a circuit. In general, compilation is a complex and time intensive task. Most of the quantum circuit compilers are inserting SWAP or similar gates (e.g.~\cite{li2019tackling, zhang2021time}) into the original circuit in order to adapt it to the underlying architecture. The resulting circuit can be up to one order of magnitude deeper than the original~\cite{zulehner2018efficient}, thus drastically limiting their utility.

Critically, the fact that quantum circuits are often formed by repeating patterns of sub-circuits inspires an opportunity to use this information for speeding up the compilation and the routing of the qubits. For example, this is the case for many arithmetic circuits which were imported from classical computing (e.g. adders, multipliers). Such circuits consist entirely of the three-qubit Toffoli gate and the single-qubit Hadamard gate. The Toffoli gate can be mapped to a cubic structure and there is a connection between the gate and colour codes~\cite{lidar2013quantum}. Moreover, larger codes can be obtained by tiling unit cells representing smaller codes~\cite{lidar2013quantum}. This work therefore proposes the utilization of this tiling technique within the framework of classical standard cell design for the efficient design of large-scale quantum circuits.

\begin{figure}[!t]
    \centering
    \includegraphics[width=1.0\columnwidth]{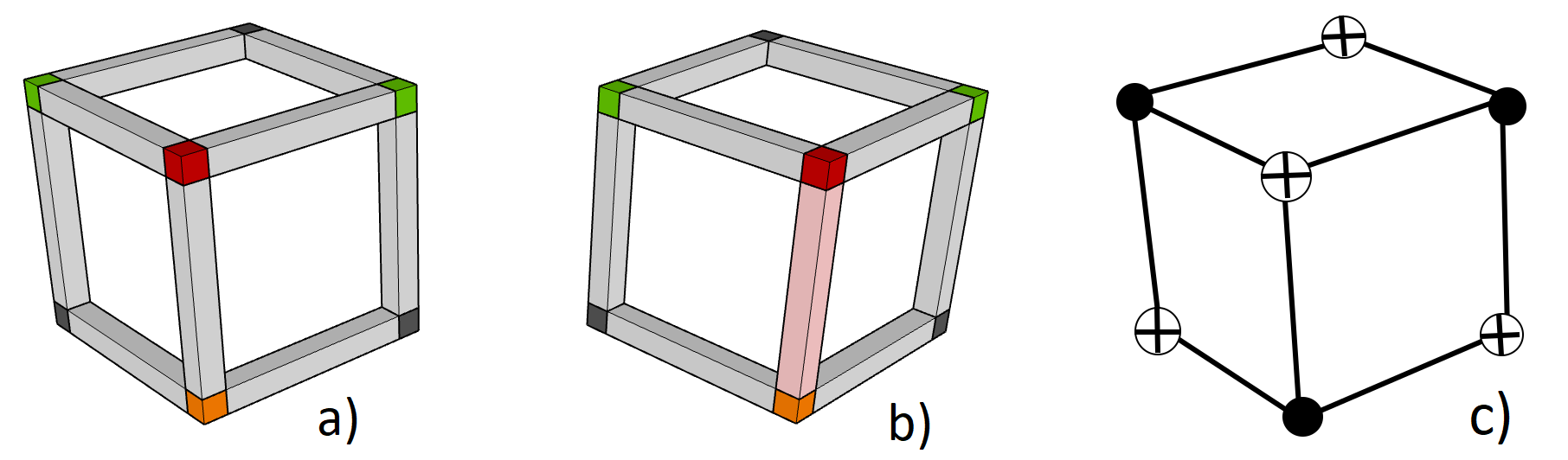}
    \caption{The standard cell for a 3D implementation of a Toffoli gate: a) Green vertices are the control qubits of the Toffoli gate, and the orange vertex is the target. In the Clifford+T decomposition of the Toffoli gate, the orange and green qubits are CNOT controls and the grey qubits are CNOT targets; b) Pink edges represent SWAPs (routing is discussed in Section~\ref{sec:res}); c) Reading a circuit from a cell is performed by replacing each vertex with a qubit, and choosing a gate that corresponds to the sticks. Assuming that T and H gates are also executed on the vertices, the circuit from (c) will correspond to Fig.~\ref{fig:tdepth1} from the Appendix. The red vertex corresponds to the qubit that is targeted by three CNOTs from the Clifford+T decomposition. These cells are not cubes, because they have only seven vertices.}
    \label{fig:cube}
\end{figure}

\subsection{Related Work}
\label{sec:rel}

The naive ``full-custom'' approach for optimizing large-scale classical circuit design automation is to automatically explore the entire range of transistor parameters and all possible interconnected circuit structures that achieve a desired complex computational function. However, this is intractable when billions or trillions of transistors are considered. Therefore, while full-custom design permits full optimization to meet design specifications, it is neither computationally nor economically viable for large system designs.

Given the extreme complexity of VLSI circuit design, the use of ``standard cells'' have therefore become a mainstream technique for the efficient design of large-scale high-performance computing systems~\cite{kahng2011vlsi} within standard circuit design curricula~\cite{grad2003standard}. In the conventional VLSI standard cell approach, a library of standard cells are developed that each contain multiple transistors with optimized parameters within an optimized circuit structure for a particular logical function. An efficient large computing system can then be automatically designed through the near-optimal interconnection of these standard cells~\cite{dunlop1985procedure}. This approach thus divides intractable large-scale design challenges into two levels of hierarchy that are both tractable, providing a computationally feasibly approach to the automated design of nearly-optimal large scale circuits.

While the standard cells themselves -- and the heuristics to design VLSI circuits with them -- are valuable proprietary intellectual property for the large chipmakers (Intel, AMD, etc.)~\cite{yang2000dragon2000}, open source options are available for academic and small-scale applications. Standard cell VLSI approaches thus enable the practical and economical design of efficient circuits with hundreds of trillions of transistors~\cite{lie2022cerebras} with a manageable sacrifice in optimality.

\subsection{Proposed Approach}

In this work, we note the fact that quantum circuit tiles are naturally analogous to standard cells, and therefore propose that large-scale quantum circuit design apply a standard cell approach adapted from classical VLSI. We particularly note that NISQ devices are tilings of various polygons such as squares (Google Sycamore), octagons and squares (Rigetti Aspen), or hexagons (IBM Hummingbird). Most of the devices are 2D, but there also exist 3D proposals such as for neutral atoms~\cite{henriet2020quantum}.

Our standard cell approach is useful for reducing the computational complexity related to the placement and routing of quantum circuits. The latter problem is related to quantum circuit design automation, and heuristics have been investigated for the past decades~\cite{maslov2008quantum, whitney2007automated} and more recently, for example, by~\cite{siraichi, li2019tackling}. Exact methods for qubit placement and routing have also been investigated, e.g.~\cite{dury2020qubo}. FPGA-like approaches to quantum circuit compilation have been proposed, for example, in~\cite{serdar2001automatic, suchara2013qure, lao2018mapping, ahsan2015optimization}. Those approaches also use cells, but those are configurable, programmable elements laid out in such a way that arbitrary computations can be implemented; similar to conventional FPGAs, such FPGA-like approaches sacrifice efficiency (speed, area, and energy) for configurability and programmability. In contrast, our cells are \emph{non-configurable, pre-programmed} elements with significant efficiency advantages. We do not have a cell for routing, like in~\cite{ahsan2015designing}, because we are extracting SWAP schedules after the circuit has been tiled.

Our approach is the opposite of the one presented in~\cite{guerreschi2018two}, in which gates are scheduled neglecting connectivity considerations, while routing operations are added at a later step. Besides classical VLSI and quantum circuits, standard cells have also been proposed for other emerging computing platforms, such as quantum-dot cellular automata~\cite{reis2016methodology, walter2019placement}, where tiles are placed and routes are computed between the tiles.

\subsection{Contributions}

\begin{figure}[!t]
    \centering
    \includegraphics[width=0.8\columnwidth]{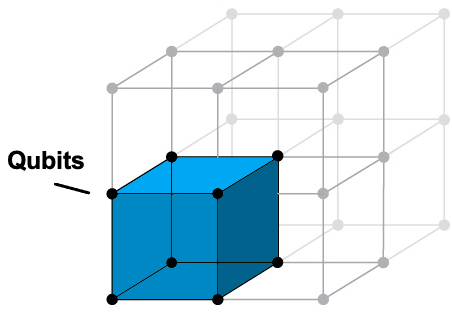}
    \caption{Quantum circuits are built from cells which are designed specifically for the underlying hardware architecture. In this example, the hardware is organised as a 3D lattice, and the standard cell (blue) might be the one illustrated in Fig.~\ref{fig:cube}.}
    \label{fig:23d}
\end{figure}

We introduce standard quantum cells, in the following also called \emph{tiles}. Our method starts from the observation that qubit lattices, as well as the quantum circuit that will be compiled, are very regular structures (e.g. Fig.~\ref{fig:23d}). We join tiles into designs that: a) represent the structure of the circuit to be compiled and, at the same time, b) are compatible with the computer's qubit arrangement.

Standard cell design approaches are suitable for the design of near-optimal quantum computing systems within reasonable design time constraints. Our work focuses on 3D lattices of qubits. Such lattices can be easily implemented by neutral atom computers, as discussed in the following section.

Our cells are designed (pre-programmed) for computations where ideally no SWAP gates are necessary within the cell. For example, no SWAPs are required in order to perform the Toffoli gate on the cell from Fig.~\ref{fig:cube}. After tiling multiple cells, qubits will still need to be routed from one tile to the other: long range interactions between remote qubits are necessary for implementing the overall computation. Consequently, our approach to compilation is the repeated execution of the following two steps:
\begin{enumerate}
    \item \emph{Routing} of qubits between the cells;
    \item \emph{Execution} of gates within the cell.
\end{enumerate}

In this paper, we express the cost of the qubit routes through the count and depth of the nearest-neighbour SWAP circuits necessary to implement the routing. Nevertheless, in the context of neutral atom computers, the SWAP cost can be replaced by the cost of shuttling the qubits.

Tiling reduces significantly the time it takes to compute routes for qubit movement. Due to the regular structure of the tiles and the tiled circuit, we can formulate hardware-aware routing algorithms. When compared with automatic routing methods, such as the ones available in Google Cirq, our algorithms are significantly faster (seconds instead of days) and save drastically on both SWAP depth and count (Section~\ref{sec:res}).

\subsection{Application to Neutral Atom Quantum Computers}
\label{sec:na}

Neutral-atom quantum computers have demonstrated the ability to operate on thousands of qubits (e.g.~\cite{manetsch2024tweezer}). These qubits are typically arranged in planar 2D layouts, while 3D configurations are feasible but more challenging to engineer. Notably, neutral-atom systems enable a technique called \emph{qubit shuttling} which allows for all-to-all connectivity for two-qubit gates under certain constraints. This effectively emulates 3D architectures even within a 2D layout.

Recent experiments with neutral-atom devices (e.g.~\cite{bluvstein2024logical}) have spurred interest in scalable compilation methods that exploit qubit shuttling to reduce reliance on SWAP gates. While functionally equivalent (e.g.~\cite{nottingham2023decomposing, schmid2023computational}), shuttling can impact circuit execution speed and fidelity. Trade-off analyses suggest that shuttling is a viable substitute for low-fidelity SWAP gates, but not yet a replacement for high-fidelity ones~\cite{schmid2023computational}.

One promising model is the dynamically field-programmable qubit array (DPQA)~\cite{tan2024compiling}, which allows for reconfigurable qubit layouts during computation. The DPQA architecture utilizes mobile traps (acusto-optic deflectors, AODs) and fixed traps (spatial light modulators, SLMs) to facilitate qubit movement between them. Two-qubit gates are performed by shuttling qubits into proximity using AODs, followed by simultaneous application of laser pulses across the entire 2D plane. This multi-step approach in DPQA mirrors our proposed \emph{route-and-execute} method. Computations within the DPQA model involve a three-step loop~\cite{schmid2023computational}: (1) loading qubits into AODs, (2) shuttling and unloading qubits into SLMs, and (3) gate execution.

Neutral-atom computers, particularly those employing AODs and SLMs, enable architectures with designated zones for specific functions~\cite{bluvstein2024logical}. These zones include memory (storing unused qubits), execution (performing gate operations), and measurement. Zone-based architectures have demonstrably facilitated pipelining in neutral-atom computations~\cite{bluvstein2024logical}. To our knowledge, our approach is the first to leverage qubit storage for co-design and compilation of circuits. This approach enables analysis and optimization of circuits as queuing networks, drawing upon prior work by Dobbs (2021).

Within the framework of specialized zones, standard cells represent the execution zones. Our approach automates the pipelining of circuit execution across sequences of zones (tiles), as detailed in Section~\ref{sec:compiling} and Figure~\ref{fig:fig4}. Furthermore, our cell design allows for the pre-computation of optimal shuttling routes, leading to more scalable compilation compared to existing methods like the hybrid-SMT approach~\cite{tan2024compiling}, hybrid heuristics~\cite{schmid2023hybrid}, and qubit mapping methods~\cite{brandhofer2021optimal}.

\section{Methods}
\label{sec:method}

We are tiling cells in order to implement larger circuits. However, a circuit's design is not the same as the circuit's execution. The latter implies the existence of a gate schedule and some qubit routing operations in order to enable the gates. However, standard cells significantly reduce the cost of scheduling and routing as we illustrate in Section~\ref{sec:res}.

We start from the following definitions. \emph{A standard cell} is a pattern that represents the 2D/3D abstraction of the qubits and the gates that form a sub-circuit (e.g. the Clifford+T decomposition of the Toffoli gate). \emph{Tiling} is the procedure by which circuits are designed in a manner that is compatible with the underlying qubit lattices. \emph{Scheduling} is the method that outputs the order in which the gates existing in the cell will be executed. For example, the goal of scheduling is to reduce the depth of the resulting circuit's gate sequence. As a result, \emph{a schedule} is a quantum circuit that includes information about gate parallelism. We \emph{extract schedules}, quantum circuits whose gates are in a well defined order, from the tiled structures.

More precisely, a standard cell is a qubit layout where the allowed two-qubit interactions are represented by sticks. For example in Fig.~\ref{fig:cube}, the grey sticks of the cell can support any two-qubit gate. However, \emph{a standard cell does not capture the order of the interactions} and this is the motivation for the necessity of extracting schedules. It should be noted that the same cell can be used to support more than a single sub-circuit. For example, a tile can be used to read NISQ circuits as well as surface code protected quantum circuits. For NISQ, the grey sticks connecting the vertices are two qubit gates (e.g. CNOT, CZ). Herein we consider that circuits are NISQy and Section~\ref{sec:res} will reflect this design decision.

For example, we can use standard cells for Toffoli gates. In Fig.~\ref{fig:cube}, coloured vertices are qubits: grey and red are ancillae, green is for Toffoli control qubits, and orange is for Toffoli target qubits. We do not explicitly mark the vertices where Hadamard gates might be applied, because these can be applied on any type of vertex. Because a qubit can change its functionality between being control or target, we will draw the same cube twice -- one next to the other.

Our cells are compatible, at the same time, with a square 2D and a cubic 3D arrangement of qubits. The three qubits of the Toffoli gate make it almost a perfect candidate for nearest neighbour 2D and 3D interactions. The AND gate from~\cite{jones2013low} is compatible only with 2D without requiring any SWAP gates. The Toffoli gate from~\cite{nielsen2010quantum} as well as the one from~\cite{gidney2018halving} are not nearest-neighbour compatible, but can be adapted to 2D with a small SWAP cost. In the following, we focus on the 3D arrangement of qubits and the corresponding tiles. We have chosen 3D because large scale quantum computers will need to operate on large numbers of qubits, and while 2D seems nowadays to be the best option due to technological constraints, 3D might be necessary in order to achieve compact quantum computers.

This work therefore describes the application of this tiling method to the design of a multiplication circuit. By creating a tile and repeatedly using it in a circuit, this standard cell approach leads to the realization of quantum circuits that are superior to those developed through sophisticated algorithms. Additionally, this standard cell approach involves a much smaller design space, and will therefore enable the automated design of large-scale quantum circuits with reasonable computational resources; this capability is not possible with conventional algorithms.

\subsection{Compiling using Standard Cells}
\label{sec:compiling}

Practical quantum circuits have a regular structure. For example, the ripple-carry adder has a V-shape of Toffoli gates, and quantum multiplication circuits are built on top of quantum addition. We use the example of a multiplication circuit, and construct manually the multiplier circuit  from~\cite{munoz2018quantum} (see Appendix for its circuit diagram) by tiling the tiles from Fig.~\ref{fig:cube}.

\begin{figure}[!h]
    \centering
    \includegraphics[width=0.8\columnwidth]{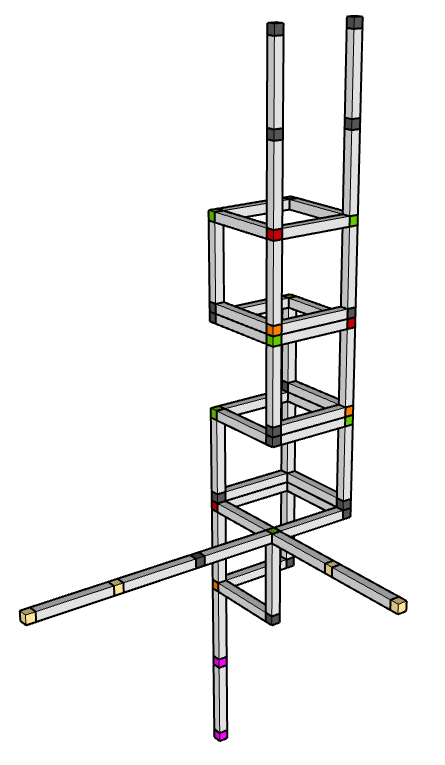}
    \caption{Four tiles (cubic standard cells) and five qubit queues (linear structures) for the 3D layout of a quantum multiplication circuit. The circuit will operate on two registers: $A$ and $B$. The lines of extending out of the figure are queues meant to hold qubits to be used within the multiplier. Specifically, the upper-left queue holds ancillae which will swap with qubits from the $B$ register, the upper right queue holds qubits from the product register, the magenta queue on the bottom holds ancillae which will eventually swap with members of the product register, and the yellow queues going out hold the $B$ operand as well as the designated ancillary qubit $Z$ (the sole grey qubit among the yellow qubits in the queue). These queues are often depicted in abbreviated forms in other figures for the sake of readability, and in fact their form is arbitrary as they will always be chains of qubits.}
    \label{fig:fig0}
\end{figure}

The resulting tiled structure includes \emph{lines} called \emph{qubit storage}, which are regions where qubits are stored whenever not used during the computation. The storage works like a queue: qubits are pushed and popped from the storage. At the same time, qubits can be moved between cells in order to be interacted with other qubits or to reach other storage. We analysed the functionality of such queues for the restricted case of a four qubit multiplication circuit in~\cite{dobbs2021fast}.

The tiling from Fig.~\ref{fig:fig0} is for a four qubit multiplier from~\cite{munoz2018quantum}. The multiplication circuit computes the product in a register that is disjoint from the multiplicand registers $A$ and $B$. The multiplier consists of: 1) a Toffoli step and 2) $n-1$ Ctrl-Add (controlled addition) steps. Some of the control signals necessary for the Ctrl-Add steps have to be available next to the Toffoli gate that is executed at a given moment. Our tiling was obtained after including some structural circuit optimisations.

\begin{figure*}[t!]
    \centering
    \includegraphics[width=\textwidth]{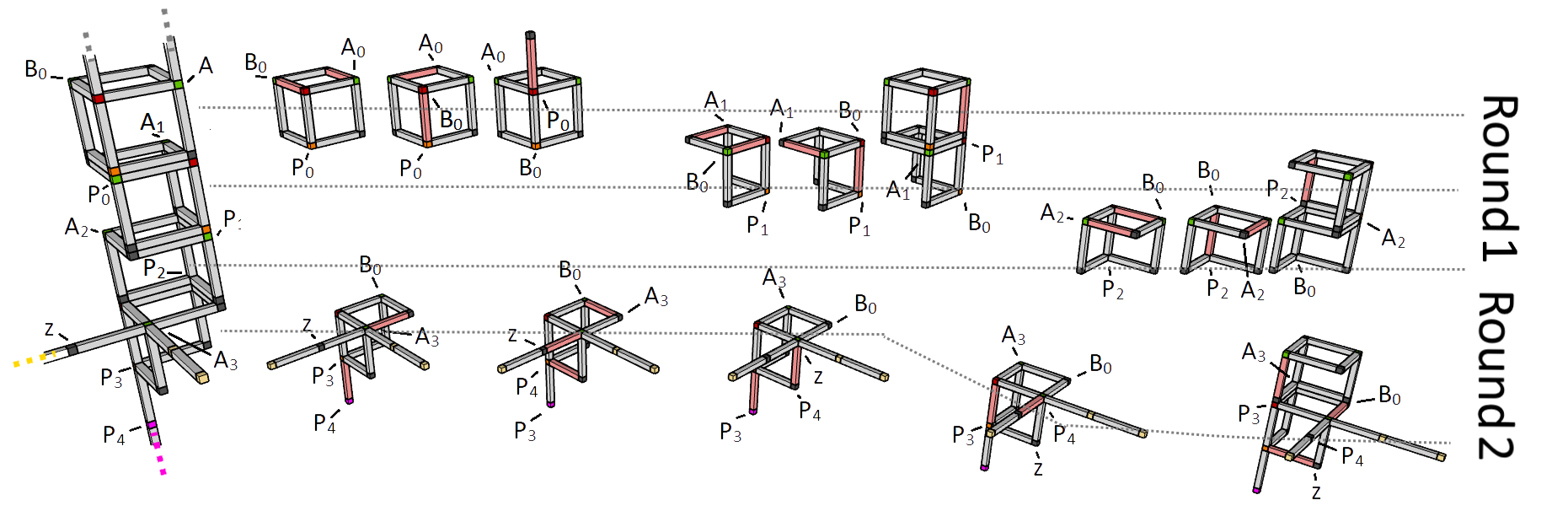}
    \caption{SWAP schedule for the Toffoli step of the 3D multiplier circuit. Red bars indicates the application of a SWAP gate between two qubits. The initial mapping of the qubit registers to be multiplied ($A$ and $B$) is indicated with labels of the form $A_i$, $B_i$. The product register is $P$, and $Z$ is an ancilla. In this step of the multiplication, a Toffoli gate is applied to each qubit $P_i$ of the product register, with the corresponding $A_i$ and $B_0$ qubit acting as controls. With respect to the scheduling procedure from Listing~\ref{lst:schedule1}, there are two rounds: Round~1 represents three repetitions of Line~\ref{line:x}, one per tile/cube. Round~2 represents Lines~\ref{line:y}-\ref{line:last}. On each row, time flows from left to right. The coloured dotted lines represent the extension of the qubit queues. The gray dotted lines indicate the position of the tile in the left hand side figure where the SWAP gates are performed. The mapping of the qubits illustrated on the left hand side figure is changing during the execution of the schedule (ie. qubits are swapped). For example, in Step~2, after being swapped, $B_0$ is close to the $Z$ ancilla.} 
    \label{fig:fig4}
\end{figure*}

The \emph{yellow storage} in Fig.~\ref{fig:fig4} holds qubits from the $B$ register to be used in later iterations (i.e. those which are used as the control qubit in each iteration), the magenta storage holds elements of the product register to be used in later iterations grey (top), and the grey storages hold elements of the product and $B$ registers which will not be used again in the computation, with the left and right storages holding $B$ register and $P$ product register qubits respectively.

\subsection{Exact Cost of Routing}
\label{sec:routing}

Standard cells are advantageous, but these still do not completely solve the need for almost arbitrary all-to-all connectivity. Nevertheless, one of the advantages of tiling is that it simplifies the calculation of the qubit routes necessary during the computation. One of the disadvantages of the current generation of quantum circuit mapping and compilation methods is their inefficiency with respect to execution times~\cite{paler2022energy}, as well as their inability to compile optimal circuits with respect to SWAP gate counts and depths.

For our tiled multiplication circuit example, the upper bound on the SWAP gate metrics (the SWAP gate count $SwapC$, and the SWAP gate depth $SWAPD$) is determined by the two types of steps: Toffoli and Ctrl-Add. For the Toffoli step which occurs once in the multiplier: 
\begin{align*}
    SwapC_t &= 5(n-1) + 12\\
	SwapD_t &= 2(n-1) + 5 
\end{align*}
	
\noindent $SwapD_t$ can be lowered to $2(n-1)$ considering that the last step can be always parallelized with the next Toffoli.

For Ctrl-Add steps we have the following: 
\begin{align*}	
	SwapC_{ca} &= 6(n-1) + 16\\
	SwapD_{ca} &= 4(n-1) + 10
\end{align*}

\noindent Again, $SwapC_{ca}$ may be lowered when considering the real CNOT cost of  the SWAPs.  The Ctrl-Add circuits swap adder qubits with ancillae in $\ket{0}$. For these operations it is sufficient to have SWAPs formed of two CNOT gates. $SwapD_{ca}$ may also be improved when considering Toffoli gate parallelism.

There are also SWAPs taking place between the Ctrl-Add steps. These SWAPs occur for $n-2$ times:
\begin{align*}
	SwapC_b &= 4(n-1) + 9\\
	SwapD_b &= 5
\end{align*}

As a result, the total number of SWAPs, not including the ones in qubit storage (their depth is 0 because they are in parallel with other SWAPs in the circuit, and their count is order of $\mathcal{O}(n^2)$):
\begin{align*}
SwapC  &= SwapC_t + (n-1)SwapC_{ca} + (n-2)SwapC_b \\
        &= 10n^2 + 6n - 13\\
SwapD  &= SwapD_t + (n-1)SwapD_{ca} + (n-2)SwapD_b \\
        &= 4n^2 + 5n - 13
\end{align*}

\subsection{Low-Depth Circuits by Layout-Aware Routing}
\label{sec:lowdepth}

We illustrate the gate scheduling of the tiled multiplication circuit. The goal is to parallelise as many gates as possible: T gates, CNOTs and SWAPs. We analyse the cost of long range interactions in terms of SWAP gate counts. In the following, we present one of the algorithms used for extracting the schedules. In the Appendix we present the algorithms for the other steps. We call our scheduler \emph{layout-aware}, because it takes into account the structure of tiles that are forming the circuit.

In the listings, ``upper" and ``lower" refer to the top and bottom of the Toffoli-cube being worked in respectively; e.g. ``lower W" refers to the ancillary qubit in the West corner of the bottom four corners of the current Toffoli-cube. ``N" is for North. Multiple gates listed on a line are in the same moment (a \emph{moment} is taken from the Cirq terminology, and represents all the gates that can be executed in parallel).

\begin{lstlisting}[escapechar=|, caption={The Toffoli step schedule.}, mathescape, label={lst:schedule1}]
01: Repeat from i = 0 to i = n-2|\label{line:x}|
02:    Toffoli (ctrl, $A_i$, $B_i$) 
03:    SWAP (ctrl, W), SWAP($A_i$, N)
04:    SWAP (ctrl, $B_i$), SWAP ($A_i$, N) 
05:    SWAP($B_i$, ancilla directly above)

06: It remains to apply the Toffoli on the final cube and clean up:
07: Toffoli(ctrl, $A_{n-1}$, $B_{n-1}$)
08: SWAP($A_{n-1}$, S), SWAP($B_{n-1}$, $B_n$) |\label{line:y}|
09: SWAP(W, z), SWAP($B_n$, lower W), 
       SWAP($A_{n-1}$, ctrl)
10: SWAP($B_n$, z), SWAP($B_{n-1}$, lower N)
11: SWAP($B_n$, empty space in queue),
       SWAP($B_{n-1}$, upper N)
12: SWAP($B_{n-1}$, N above current position),
       SWAP(z, lower N), SWAP(ctrl, W)
13: SWAP(ctrl, next ancilla (B) in queue))|\label{line:last}|
\end{lstlisting}

Throughout the schedule, gates are typically applied to either some combination of $ctrl$, $z$, $A_j$, and $p_j$, where j is the largest subscript in the current iteration, or to each of $A_i$ and $B_i$ for every cube i, where $a$ and $b$ refer to the registers of the first and second operands of the current controlled addition step respectively (in the case of the Toffoli step, these refer to the operands of the first controlled-addition step). In the former case, qubits must be swapped around the very small space in the final cube, while in the latter case $ctrl$ is swapped up the tower to be used in each cube.

Fig.~\ref{fig:fig4} (and Figs.~\ref{fig:fig2},~\ref{fig:fig3} and~\ref{fig:fig1} from the Appendix) demonstrate visually each moment of SWAPs during the circuit, where a red bar between two qubits represents a SWAP gate being applied between them. Fig.~\ref{fig:fig4} visualises each step of SWAPs necessary to implement the Toffoli step of the sample 4-qubit multiplier. Note that these figures only represent SWAP moments, so the application of Toffoli gates is not shown (though they are present in the schedules below). We have chosen to not introduce additional qubits and to perform SWAPs only across the multiplication tiling. For example, the number of SWAPs could be improved by adding a \emph{vertical tower} parallel to the existing structure.

\begin{figure*}[!t]
    \centering
    \includegraphics[width=0.4\linewidth]{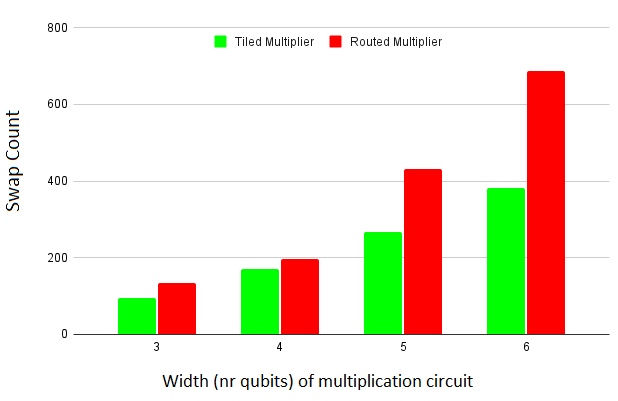}
    \hfill
    \includegraphics[width=0.4\linewidth]{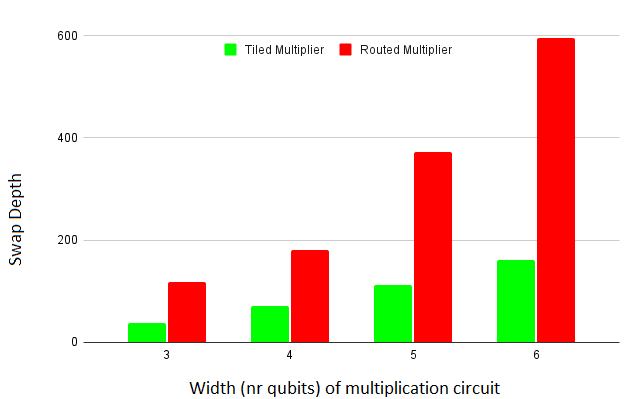}
    \caption{Automatically routed circuits vs. tiled circuits -- SWAP depth and count for different multiplication circuit sizes. The left chart compares SWAP counts and the right chart SWAP Depth. The horizontal axis in both charts represents the size of the multiplication (e.g a 4-bit multiplication), and the vertical axis is the number of SWAP gates and the number of SWAP moments required respectively. The green bars are our tiled multiplier, whereas the red bars are the automatic Google Cirq routing. We conclude that tiling, for these kind of circuits, is more resource efficient.}
    \label{fig:chart_fig}
\end{figure*}

\section{Results: Fast Compilation, Scalable and Efficient Routing}
\label{sec:res}

The results from this manual tiling are compared to automatically mapped and routed multiplication circuits from~\cite{munoz2018quantum}. We ran our experiments on an 8th gen i7 processor with 16GB of RAM. We are using Google Cirq, although there exist alternative compilers, because: a) our circuits were written in Cirq (see source code at~\cite{gitrepo}); b) already had our own scheduler which outperforms any automatic compiler in terms of execution speed -- other compilers would have timed out at larger multiplication circuits, but nevertheless could not have kept pace with the architecture-aware scheduler; c) the scheduler returns highly parallel SWAP networks which are difficult to obtain automatically. Consequently, the choice of the compiler would not have played an important role in highlighting the scalability of our method.

Fig.~\ref{fig:chart_fig} compares the SWAP counts and SWAP depths of the circuits compiled with Cirq and our tiled versions. When compared with automatic routing methods, such as the ones available in Google Cirq, our tiled version of the multiplier improves significantly on both SWAP depth and count.

Fig.~\ref{fig:chart_fig} illustrates the comparison between the counts and depths of the manually tiled circuit and the routing costs computed automatically using Cirq. The SWAP depth and count numbers for Cirq were obtained by providing the Cirq router with a 3D device of the same general dimensions as our tiled circuit and allowing it to route greedily. This is an extremely challenging problem, because the router needs to map the circuit qubits to the device qubits in a way that allows for non-overlapping routes in an efficient manner (lower depth). To simplify the mapping and the routing, in many cases the router was actually provided with more qubits than the tiled circuit uses. For additional details and source code see the repository at~\cite{gitrepo}.

The compilation of the 6-qubit multiplier required more than 3 days. We repeated the compilation multiple times and measured the same duration. We did not succeed to compile using Cirq 7-qubit wide multipliers automatically within reasonable time (one week).

We define the \emph{usage ratio} as the tile qubits to the total number of computer qubits necessary to hold the tiling. In the case of a single 3D tile we use seven qubits, and if the hardware would look like a cube (has eight vertices), the \emph{usage ratio} would be 7/8. If one considers that only three are computational qubits and four are ancilla, then the effectiveness ratio is 3/8. Another perspective is that four qubits are needed in order to achieve high parallelism in a 3D layout.

A more detailed analysis of the usage and effectiveness ratios for the multiplier shows that, after some optimization on the positioning and structure of qubit storages, the multiplier can achieve a usage ratio of 33/48 for the 4-qubit case. Generalized, the multiplier requires $2 * 3 * (N+1 + \left \lceil{N / 4}\right \rceil + \left \lfloor{N/2}\right \rfloor)$ qubits, and uses some number less than this. Note that the portion of the above equation in parentheses represents the height of the multiplier structure, whereas the width and length remain constant as 2 and 3. This information is relevant as it provides for a highly regular final structure for a quantum arithmetic circuit, which is likely to be repeated many times throughout a given computation.

\section{Conclusion}
\label{sec:conclusion}

Tiling is a method for compiling circuits for a device that has a regular layout of qubits, and can be used to improve the usage ratio of the quantum chips as a whole. Standard cells and tiling, together with layout-aware routing methods, allow for the extremely fast and efficient compilation of very large scale quantum circuits.

Tiling is especially useful for highly regular, frequently repeated sub-circuits such as those in quantum arithmetic. We illustrated the capabilities of standard cells by the example of using 3D tiles which were specifically designed for cubic qubit lattices. We demonstrated the effectiveness of our method by using it to design a quantum multiplier, showing its usefulness for scheduling SWAP gates by significantly improving on both the SWAP depth and count of the circuit over existing automatic routing methods. 

We conclude that tiling standard cells allows for a faster and improved understanding of the layout of the compiled circuit without the processing time involved in compilation and routing. It is a valuable tool in estimating the resources required for compiling a given quantum circuit to hardware, and especially in creating structures which have highly parallel SWAP schedules.

There exist multiple applications of standard cells and tiling. First, tiling quantum circuits can inform the co-design of computing architectures, where the qubit layout, for example, is developed in parallel to the circuits to execute. Such 2D and 3D architectural co-design can be implemented with neutral atoms~\cite{henriet2020quantum}, for example. In neutral atom computers, qubits can be easily shuttled and moved between specialized zones (memory, execution, measurement). Standard cells are analogous to execution zones, and by tiling cells one is designing a pipelined architecture, where qubits are moved/shuttled between zones. This allows to determine optimal shuttling schedules for the movement of the qubits between the zones.

Second, the relation between standard cells and quantum error-correcting codes and quantum error mitigation techniques, e.g.~\cite{chao2018quantum}, make tiling an interesting candidate for the co-design of fault-tolerant circuits.

Third, tiling is extremely useful for performing quantum resource estimation, which is concerned with determining the number of qubits and the time necessary to run a computation. In order to achieve large-scale quantum computations, the resources required for implementing fault-tolerance and error-correction will have to be reduced significantly (e.g. the error-corrected implementation of Shor's algorithm would require 20 million qubits\cite{gidney2021factor}). Fast resource estimation can be performed by combining standard cells into larger structures. In doing so, resource estimation becomes a bottom-up approach: at the lower level the cells are optimized, and the higher level the cells are tiled and resource estimation is also simplified by using the SWAP schedulers which are faster than conventional quantum circuit compilers.

In future work, the superiority of our proposed tiling approach will translate to drastic improvements in the automation of large-scale quantum circuit design. To rapidly advance this tiling approach, we will leverage the enormous literature on VLSI standard cell designs and the experiences gathered through decades of effort by the microelectronics industry. Future work will also focus on tiling as a means to reduce the number of qubits which are inaccessible in large, error-corrected computations by virtue of the regular structure of the models it creates.

\section{Acknowledgments}
This research was developed in part with funding from the Defense Advanced Research Projects Agency [under the Quantum Benchmarking
(QB) program under award no. HR00112230007 and HR001121S0026 contracts]. We thank George Watkins for feedback on a preliminary version of this manuscript.

\bibliographystyle{IEEEtran}
\bibliography{__main}

\section*{Appendix}

\begin{figure}[!h]
    \centering
    \includegraphics[width=\columnwidth]{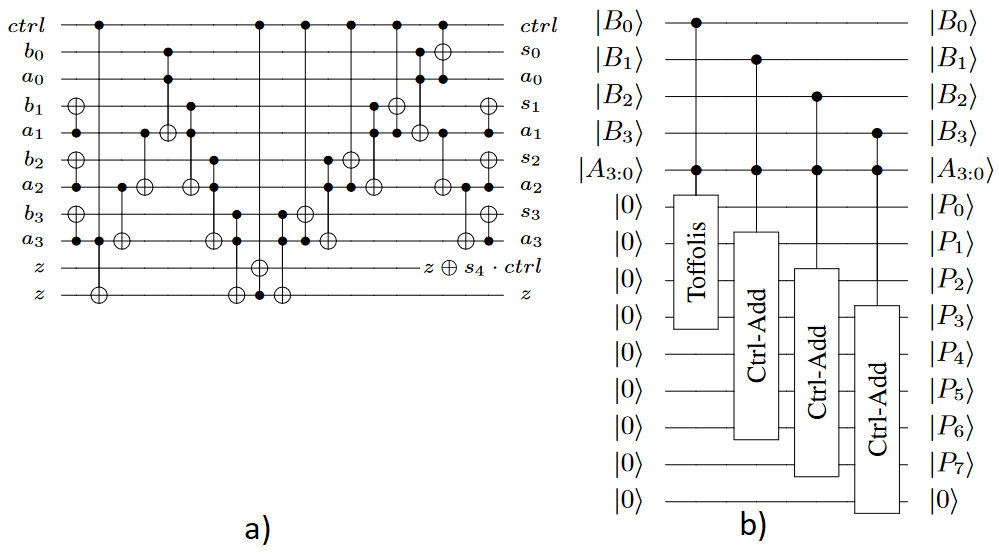}
    \caption{Circuit from~\cite{munoz2018quantum} which we used to construct tilings: a) the controlled-adder; b) the multiplier which is a sequence of controlled-adders.}
    \label{fig:munoz}
\end{figure}

\begin{figure}[!t]
    \centering
    \includegraphics[width=\columnwidth]{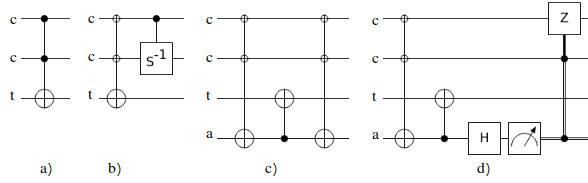}
    \caption{The Toffoli gate from a) can be implemented using: b) an AND gate and a controlled-S gate, c) an ancilla initialised in $\ket{0}$, two AND gates and a CNOT, d) an AND, an ancilla, and a CNOT; here the uncomputation is measurement-based. The wire labels denote [c]ontrol, [t]arget, and [a]ncilla.}
    \label{fig:toff_and}
\end{figure}

Our tiles are useful, for example, for Toffoli+H circuits. Toffoli gates cannot, in general, be executed natively, and are decomposed into sequences of Clifford+T gates. The latter are compatible with the NISQ device or the error-correcting code. The Hadamard gate is comparatively straightforward to execute on almost any computer. A Toffoli gate can be obtained from an AND gate and a controlled-S gate. When using measurement-based uncomputation~\cite{jones2013low} and an ancilla, the Toffoli gate can be implemented with a single AND gate. These circuits are illustrated in Fig.~\ref{fig:toff_and}. The circuit diagram for the ripple-carry adder and the multiplication circuit using it are illustrated in Fig.~\ref{fig:munoz}.

\begin{figure}[!h]
    \centering
    \includegraphics[width=0.8\columnwidth]{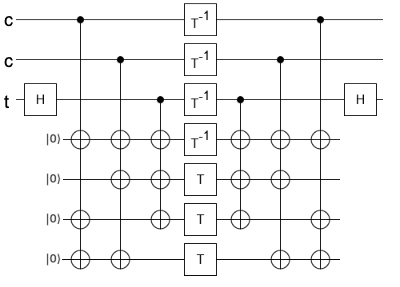}
    \caption{The T-depth 1 CCZ gate decomposition~\cite{selinger2013quantum}.}
    \label{fig:tdepth1}
\end{figure}

\begin{figure*}
    \centering
    \includegraphics[width=0.9\textwidth]{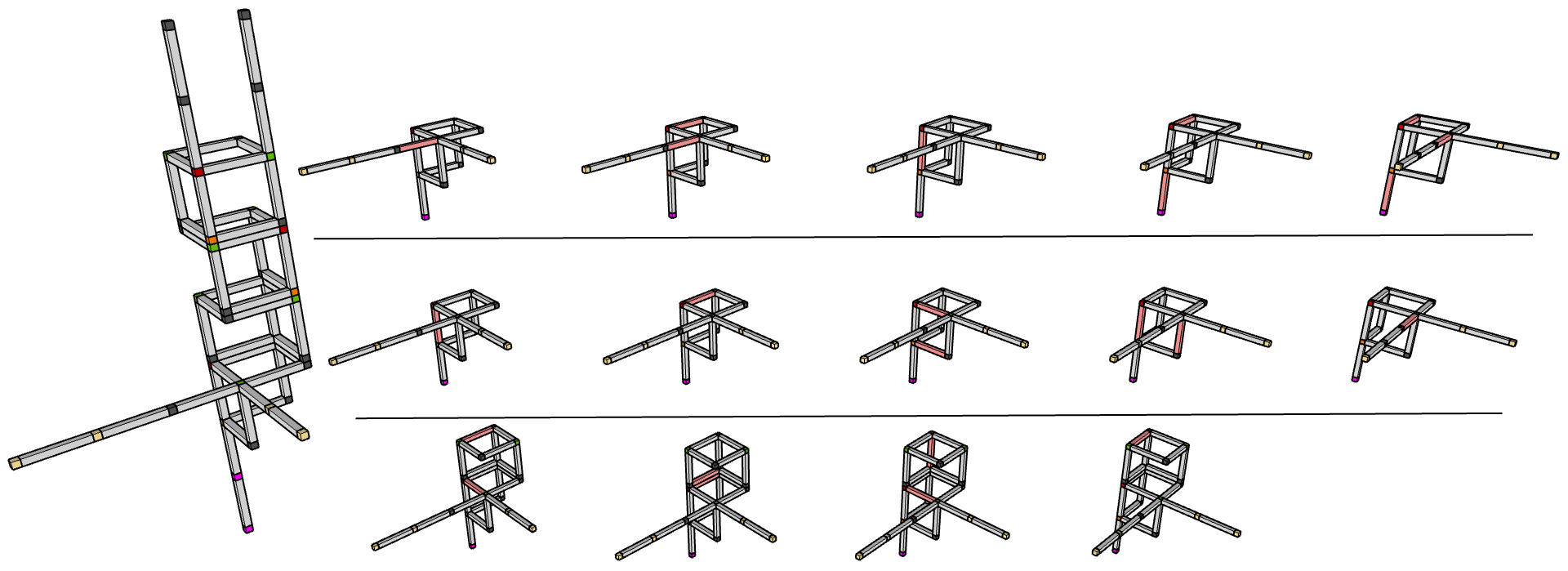}
    \caption{Controlled-addition Step - SWAP schedule for the first part of the controlled-addition step of the 3D multiplier circuit, where a red bar indicates the application of a SWAP gate between two qubits. Time flows from left to right one row at a time. The logical operations to be performed on the circuit are applied between SWAP gates, as described in the schedule. By the end of one iteration of this step, one additional product qubit has been computed. Following the reset step, this step will be repeated for another iteration, and so on for N-1 iterations.}
    \label{fig:fig2}
\end{figure*}

\begin{figure*}
    \centering
    \includegraphics[width=0.9\textwidth]{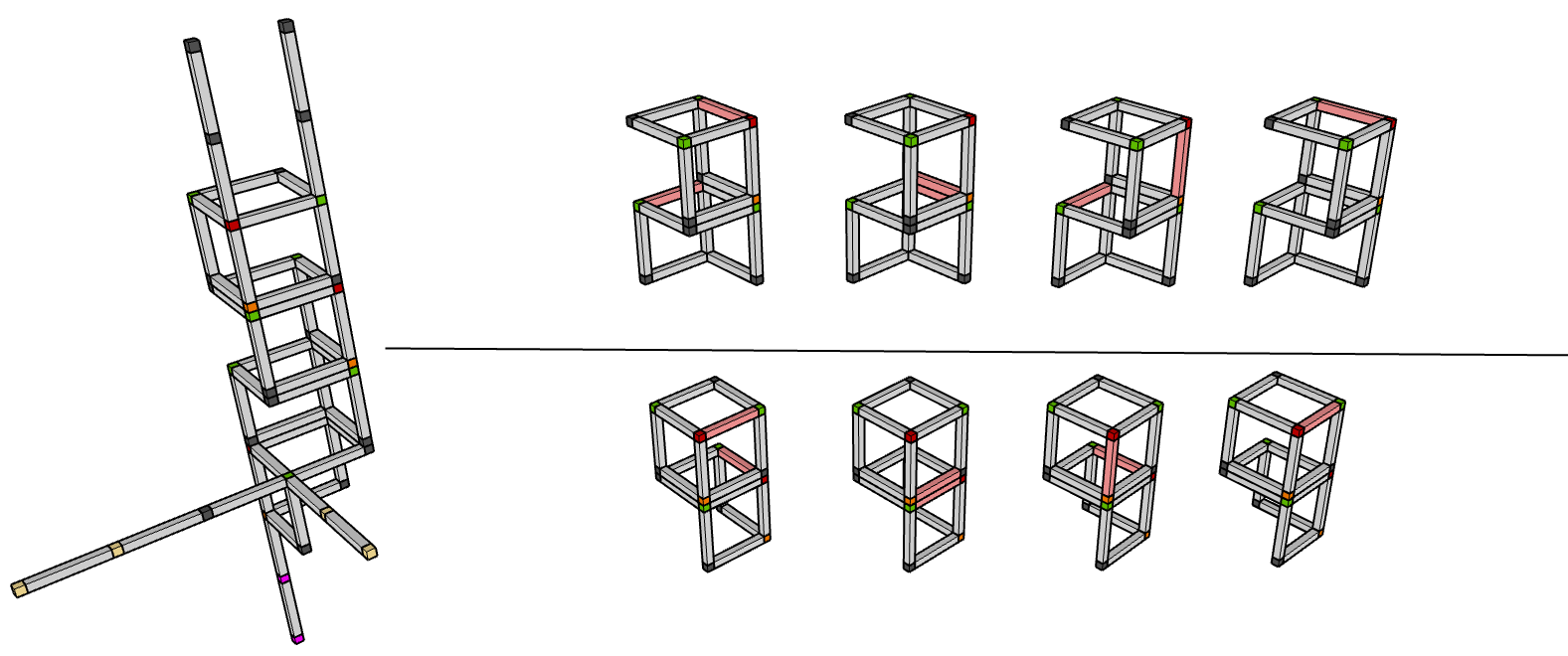}
    \caption{Controlled-addition Step - SWAP schedule for the second part of the controlled-addition step of the 3D multiplier circuit, where a red bar indicates the application of a SWAP gate between two qubits. Time flows from left to right, one row at a time.}
    \label{fig:fig3}
\end{figure*}

\begin{figure*}
    \centering
    \includegraphics[width=0.9\textwidth]{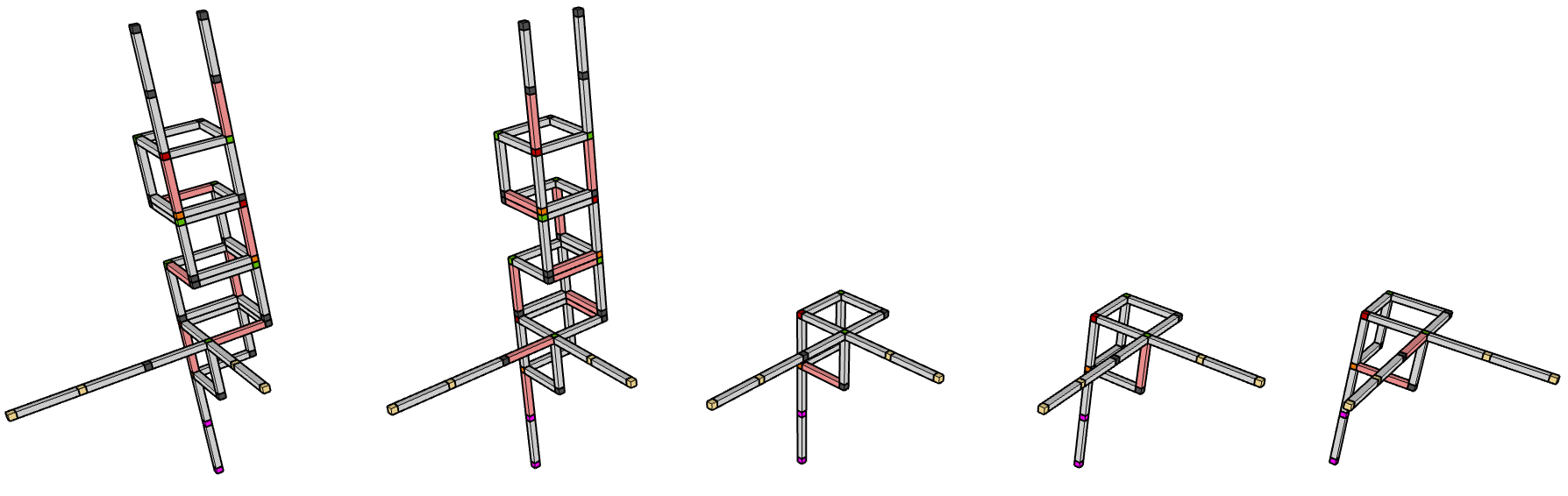}
    \caption{Reset Step - SWAP schedule for the reset step of the 3D multiplier circuit, where a red bar indicates the application of a SWAP gate between two qubits. This step is applied following each iteration of the controlled-addition step. During this step, the control qubit ($B_1$ in the first iteration) is replaced by the next qubit in the B register. Time flows from left to right, one row at a time. Regardless of the size of the multiplication, this step always has a SWAP depth of 5.}
    \label{fig:fig1}
\end{figure*}

We present two other schedules for the Ctrl-Add and the third step of the multiplication circuit. Figs. ~\ref{fig:fig2} and ~\ref{fig:fig3} are the visualisation of the corresponding schedules of a single controlled-addition iteration, and Fig.~\ref{fig:fig1} demonstrates the SWAP moments necessary after each iteration of controlled-addition. 

\begin{lstlisting}[escapechar=|, caption={The jth iteration of the Ctrl-Add subcircuit, which repeats $N-1$ times in the multiplier circuit, running a setup subcircuit between each repetition after the first. Note that ctrl = $B_j$.}, mathescape, commentstyle=\color{blue}\ttfamily, language=C, label={lst:schedule2}]
01: Toffoli(ctrl, $A_3$, z)
02: Toffoli($B_j$, $A_0$, $A_1$)
03: Toffoli($B_{j+1}$, $A_1$, $A_2$)
04: Toffoli($B_{j+2}$, $A_2$, $A_3$)
05: SWAP($B_{j+3}$, ctrl)
06: Toffoli($B_{j+3}$, $A_3$, z)
07: SWAP($B_{j+3}$, ctrl), SWAP($A_3$, N)
08: SWAP($A_3$, z)
09: SWAP(z, E), SWAP($A_3$, $B_{j+4}$)
10: Toffoli(ctrl, z, $B_{j+4}$)
11: SWAP($B_{j+3}$, ctrl), 
        SWAP($A_3$, $B_{j+4}$), SWAP(z, N)
12: SWAP(z, $A_3$)
13: SWAP($A_3$, E)
14: Toffoli($B_{j+3}$, $A_3$, z)
15: SWAP(z, lower W)
16: SWAP($B_{j+3}$, z) 
17: SWAP(z, ctrl), SWAP($B_{j+3}$, lower N)
18: for(k = n-1, k >=0; k--)
19:     Toffoli(ctrl, $A_k$, $B_{j+k}$)
20:     Toffoli($B_{j+k-1}$, $A_k-1$, $A_k$)
21:     SWAP(ctrl, next clockwise ancilla), 
           SWAP($B_{j+k-1}$, next clockwise ancilla)
22:     SWAP(ctrl, $A_k$)
23:     SWAP(ctrl, $B_{j+k-1}$),
           SWAP($A_k$, next counter-clockwise ancilla)
24:     SWAP(ctrl, next counter-clockwise ancilla)
25: Toffoli(ctrl, $A_0$, $B_j$)
\end{lstlisting}

\begin{lstlisting}[escapechar=|, caption={The SWAP schedule immediately following the $j$-th iteration of the Ctrl-Add subcircuit.}, mathescape, commentstyle=\color{blue}\ttfamily, language=C, label={lst:schedule3}]
01:  SWAP(ctrl, queue above)
02:  SWAP($B_{lowest}$, queue above)
03:  for each cube k (besides the first):
04:     SWAP($A_k$, next counter-clockwise ancilla)
            SWAP($B_{j+k}$, $ancillA_{above}$)
05:  for each cube k (besides the first):
06:     SWAP($A_k$, next counter-clockwise ancilla), 
            SWAP($B_{j+k}$, $ancillA_{above}$), 
            SWAP(z, ancilla in control position)
07:  SWAP($B_{j+n}$, N)
08:  SWAP($B_{j+n}$, W)
09:  SWAP($B_{j+n}$, Z)
10:  SWAP(Z, N), SWAP($B_{j+n}$, B_${j+1}$)
\end{lstlisting}

\end{document}